# Review Paper:

# Hybridization effect in coupled metamaterials

Hui LIU(刘辉), Tao LI(李涛), Shu-ming WANG(王漱明) and Shi-ning ZHU(祝世宁)

*National Laboratory of Solid State Microstructures, Department of Physics, Nanjing University, People's Republic of China*

Email: liuhui@nju.edu.cn, zhusn@nju.edu.cn; URL: http://dsl.nju.edu.cn/mpp

**Abstract:** Although the invention of the metamaterials has stimulated the interest of many researchers and possesses many important applications, the basic design idea is very simple: composing effective media from many small structured elements and controlling its artificial EM properties. According to the effective-media model, the coupling interactions between the elements in metamaterials are somewhat ignored; therefore, the effective properties of metamaterials can be viewed as the "averaged effect" of the resonance property of the individual elements. However, the coupling interaction between elements should always exist when they are arranged into metamaterials. Sometimes, especially when the elements are very close, this coupling effect is not negligible and will have a substantial effect on the metamaterials' properties. In recent years, it has been shown that the interaction between resonance elements in metamaterials could lead to some novel phenomena and interesting applications that do not exist in conventional uncoupled metamaterials. In this paper, we will give a review of these recent developments in coupled metamaterials. For the "meta-molecule" composed of several identical resonators, the coupling between these units produces multiple discrete resonance modes due to hybridization. In the case of a "meta-crystal" comprising an infinite number of resonators, these multiple discrete resonances can be extended to form a continuous frequency band by strong coupling. This kind of broadband and tunable coupled metamaterial may have interesting applications. Many novel metamaterials and nanophotonic devices could be developed from coupled resonator systems in the future.

I.   Introduction
II.  Hybrid modes introduced by coupling effect in meta-molecules
III. Coherent collective waves in one-dimensional meta-chains
IV.  Excitation of coupled plasmon modes in two-dimensional meta-crystals
V.   Outlook

## I. Introduction

All classical electromagnetic (EM) phenomena in various media are determined by the well-known Maxwell's equations. To describe the EM properties of a material, two important parameters are introduced, that is, electric permittivity $\varepsilon$ and magnetic permeability $\mu$. In principle, if the $\varepsilon$ and of materials are known, then the propagation of EM waves inside materials, or the EM phenomena at the surface between two materials can be well predicted. For example, the refraction of an EM wave at the interface is described by Snell's law, $\sin\theta_i / \sin\theta_r = n_r / n_i$, which states that the relation between the incident angle ($\theta_i$) and the refracted angle ($\theta_r$) is determined by the refractive index, $n = \sqrt{\varepsilon\mu}$, of the two media involved.

Clearly, if we can modify $\varepsilon$ and $\mu$ artificially, then the propagation behavior of EM waves in the material can be manipulated at will. For instance, in 1967 when Veselago first theoretically studied the EM properties of a material with a negative refractive index (simultaneously negative $\varepsilon$ and $\mu$), he found that light will be refracted negatively at the interface between such a material and a normal positive index material [1]. This so-called "negative refraction" phenomenon does not violate the laws of physics, yet it challenges our physical perception and intuition. In such negative index media (NIM), a number of other surprising phenomena were also predicted, such as the reversed Doppler shift and Cerenkov radiation. However, Veselago's work was ignored for a long time because no such double negative materials (i.e., where both $\varepsilon < 0$ and $\mu < 0$) are obtainable in nature, making negative refraction seemingly impossible. Indeed, we are limited by the natural material properties. Most dielectrics only have positive permittivities. For most metals, $\varepsilon < 0$ can be met at optical range, and the plasma frequency can be moved downwards into microwave range by replacing the bulk metal with a rodded medium [2-4], yet permeability is always positive. Negative $\mu$ is accessible in some ferromagnetic materials in the microwave region, but they are difficult to find above terahertz frequencies in the natural world.

In recent years, to achieve designable EM properties, especially negative $\mu$ at high frequencies, people have invented novel artificial materials known as metamaterials. The basic idea of a metamaterial is to design artificial elements that possess electric or magnetic responses to EM waves. Many such elements can work as artificial "atoms" to constitute a metamaterial "crystal." The geometric size of these atoms and the distances between them are much smaller than the wavelength of EM waves. Then, for an EM wave, the underlying metamaterial can be regarded as a continuous "effective medium." Correspondingly, the property of a metamaterial can be described by two effective parameters: effective permittivity $\varepsilon_{eff}$ and permeability $\mu_{eff}$. In 1999, Pendry first designed a metallic magnetic resonance element: a split-ring resonator (SRR) [5]. When an SRR is illuminated by light, the magnetic component of the EM wave induces the faradic current in this structure, giving rise to a magnetic dipole. Using SRRs as structure elements, Pendry constructed a new kind of magnetic metamaterial. The effective permeability, $\mu_{eff}$, of this metamaterial has the form

$$\mu_{eff} = 1 - \frac{F\omega^2}{\omega^2 - \omega_{mp}^2 + i\gamma\omega} \qquad (1)$$

where F is the fractional volume of the cell occupied by the SRR. Equation (1) suggests that $\mu_{eff}$ follows a Drude-Lorentz resonance, and $\mu_{eff}$ can be negative around the frequency $\omega_{mp}$ if the damping term $\gamma$ is not so large. Such plasmon resonance in SRR is caused by a magnetic field, so the corresponding resonance frequency, $\omega_{mp}$, here is called the magnetic plasmon (MP) frequency. Motivated by Pendry's work, D. R. Smith combined SRR and metallic wires to construct a metamaterial with simultaneously negative $\varepsilon_{eff}$ and $\mu_{eff}$ [6]. The negative refraction proposed by Veselago was finally experimentally verified in the microwave region [7].

One of the most important applications of NIMs is a superlens, which allows imaging resolution beyond the diffraction limit [8-13]. Considering its significant applications in the visible region, increasing the MP frequency, $\omega_{mp}$, to obtain negative refraction for visible light is a very valuable and challenging task. Given that $\omega_{mp}$ arises from an inductor-capacitor circuit (LC) resonance in SRRs and is determined by the geometric size of this structure, it can be increased by shrinking the size of the SRRs. In 2004, X. Zhang and colleagues fabricated a planar structure composed of SRRs. The size of the SRR was just a few micrometers, and $\omega_{mp}$ was around 1 THz [14]. Immediately after X. Zhang's work, Soukoulis and colleagues fabricated an SRR sample with a unit cell of several hundred nanometers, and $\omega_{mp}$ was raised to 100 THz [15]. Another result is obtained around 1.5 μm, which is the telecommunication wavelength in the infrared range [16]. As the structure of an SRR is so complex, it is very difficult to decrease its geometric size any further with the existing nanofabrication technique. To obtain MP resonance at higher frequencies, people began to seek other simple MP structures. In fact, inductive coupled rod pairs are very simple structures that Zheludev and colleagues proposed as constituting chiral metamaterials [17]. Shalaev and colleagues found that such nanorod pairs could also be used to produce MP resonance and negative refraction at the optical communication wavelength of 1.5 μm [18, 19]. Almost at the same time, S. Zhang et al. proposed a double-fishnet structure to obtain negative refraction at about 2 μm [20]. Although Shalaev and S. Zhang verified that their structures possessed a negative refraction index by measuring the phase difference of the transmitted waves, they could not directly observe the negative refraction in their monolayer metamaterial structures. Until quite recently, direct negative refraction was observed by X. Zhang and colleagues in the three-dimensional bulk metamaterials of nanowires [21] and fishnet structures [22] in the optical region. Besides the aforementioned important works on negative refraction, many other studies from recent years provide a good introduction to the rapid progress that has taken place in this field [23-27]. In addition to negative refractions, MP resonance has also been applied to another metamaterial that has attracted considerable attention, namely, cloaking materials [28-31].

Although the invention of the metamaterial has stimulated the interest of many researchers and its various applications have been widely discussed, the basic design idea is very simple: composing effective media from many small structured elements and controlling its artificial EM properties. According to the effective-media model, the coupling interactions between the elements in metamaterials are somewhat ignored; therefore, the effective properties of metamaterials can be viewed as the "averaged effect" of the resonance property of the individual elements. However, the coupling interaction between elements should always exist when they are arranged into metamaterials. Sometimes, especially when the elements are very close, this coupling effect is not negligible and will have a substantial effect on the metamaterial's properties. Under such circumstances, the uncoupling model is no longer valid, and the effective properties of the metamaterial cannot be regarded as the outcome of the averaged effect of a single element (see Figure 1). Many new questions arise: How do we model the coupling in metamaterials? What new phenomena will be introduced by this coupling effect? Can we find any new interesting applications in these coupled systems?

Coupled metamaterials consisting of resonance elements with a strong coupling interaction have already developed into an important branch of metamaterial research. The "hybridization effect" caused by these coupling interactions between resonators in metamaterials is attracting increased interest. Some multiple hybrid modes or continuum collective hybrid modes were found in metamaterials after including this hybridization effect (see Figure 1). Quite a number of papers have already reported this new kind of coupled resonance modes. Various novel phenomena and properties have been explored, and these have led to many new interesting applications that do not exist in uncoupled metamaterials. In this review paper, we will give an overall introduction to these recent developments in hybrid modes that were introduced by the coupling effect in metamaterials. First, the hybridization effect of plasmonic resonance in meta-molecules is presented in section two. Second, we will describe coherent collective waves in one-dimensional meta-chains in section 3. Then, the excitation of coupled plasmon modes in a two-dimensional (2-D) system will be discussed in section four. In the last section, an outlook will be presented to predict possible future developments of hybridization effect in coupled metamaterials.

**II. Hybrid modes introduced by coupling effect in meta-molecules**

In 2003 [32], Halas and colleagues introduced a hybridization model to describe the plasmon response of complex nanostructures. It was shown that the resonance modes of a complex metallic nanosized system could be understood as the interaction or hybridization result of the elementary geometries. The hybridization principle provides a simple conceptual approach to designing nanostructures with desired plasmon resonances. In their following work, this method was successfully used to describe the plasmon resonance in a nanoshell [33], nanoparticle dimers [34], nanoshell dimers [35, 36], and nanoparticles near metallic surfaces [37] .

In fact, the hybridization model could also be applied to deal with the EM wave response of metamaterials that comprise many resonance elements. SRR is the best-known magnetic "atom" of metamaterials. Therefore, the investigation of how SRRs interact with each other is both a fundamental and typical study. Apparently, a magnetic dimer (MD) made of two SRRs is the simplest system with which to study the coupling effect [38]. In Figure 2, we present the general configuration of an MD, composed of two identical SRRs separated by a finite distance, D. To study the magnetic response of this MD, an MP hybridization model was established. In our approach, we use the Lagrangian formalism, first calculating the magnetic energy of a single SRR and later expanding the theory for a system of two coupled SRRs. For simplicity, in the analysis we consider each SRR an ideal LC circuit composed of a magnetic loop (the metal ring) with inductance L and a capacitor with capacitance C (corresponding to the gap). The resonance frequency of the structure is given by $\omega_0 = 1/\sqrt{LC}$, and the magnetic moment of the SRR originates from the oscillatory behavior of the currents induced in the resonator. If we define the total charge, Q, accumulated in the slit as a generalized coordinate, the Lagrangian corresponding to a single SRR is written as $\mathcal{L} = L\dot{Q}^2/2 - Q^2/2C$, where $\dot{Q}$ is the induced current, $L\dot{Q}^2/2$ relates to the kinetic energy of the oscillations, and $Q^2/2C = L\omega_0^2 Q^2/2$ is the electrostatic energy stored in the SRR's gap. Similarly, the Lagrangian that describes the MD is a sum of the individual SRR contributions with an additional interaction term

$$\mathcal{L} = \frac{L}{2}\left(\dot{Q}_1^2 - \omega_0^2 Q_1^2\right) + \frac{L}{2}\left(\dot{Q}_2^2 - \omega_0^2 Q_2^2\right) + M\dot{Q}_1\dot{Q}_2 \qquad (2)$$

where $Q_i$ (i = 1, 2) are the oscillatory charges and M is the mutual inductance. By substituting $\mathcal{L}$ in the Euler Lagrange equations

$$\frac{d}{dt}\left(\frac{\partial \mathcal{L}}{\partial \dot{Q}_i}\right) - \frac{\partial \mathcal{L}}{\partial Q_i} = 0 \quad (i = 1, 2) \tag{3}$$

it is straightforward to obtain the magnetic plasmon eigenfrequencies, $\omega_{+/-} = \omega_0 / \sqrt{1 \mp \kappa}$, where $\kappa = M/L$ is a coupling coefficient. The high energy or anti-bonding mode, $|\omega_+\rangle$, is characterized by anti-symmetric charge distribution ($Q_1 = -Q_2$), while the opposite is true for the bonding or low energy $|\omega_-\rangle$ magnetic resonance ($Q_1 = Q_2$). Naturally, the frequency split $\Delta\omega = \omega_+ - \omega_- \approx \kappa\omega_0$ is proportional to the coupling strength. The hybridization of the magnetic response in the case of a dimer is mainly due to inductive coupling between the SRRs. If each SRR is regarded as a quasi-atom, then the MD can be viewed as a hydrogen-like quasi-molecule with energy levels $\omega_-$ and $\omega_+$ originating from the hybridization of the original (decoupled) state, $\omega_0$. The specific nature of the MP eigenmodes is studied in Figure 3 in which the local magnetic field distributions are depicted for the low energy ($\omega_-$) and high energy ($\omega_+$) states respectively. In accordance with the prediction based on the Lagrangian approach, the SRRs oscillate in-phase for the bonding mode $|\omega_-\rangle$ and out of phase for the anti-bonding mode $|\omega_+\rangle$. Since the mutual inductance M decreases dramatically with distance, a strong change in the resonance frequencies $\omega_\pm$ is expected. This phenomenon is demonstrated in Figure 3(c-d), where the MP eigenfrequencies $\omega_\pm$ and the frequency change $\Delta\omega = \omega_+ - \omega_-$ are calculated. With a decreasing separation between the SRR, an increase in the frequency gap $\Delta\omega$ is observed. The opposite effect takes place at large distances where the magnetic response is decoupled. This result has already been experimentally proven in the microwave range [39]. The radiation property of such two coupled SRR was also investigated [40] and it found the radiation loss was greatly suppressed at the lower frequency bonding mode and was enhanced at the higher frequency antibonding mode. As the distances between the SRRs decreased, the magnitude of the radiation power of the bonding mode correspondingly decreased due to the greater cancellation of the electric dipoles. Furthermore, this suppression of the radiation loss led to a higher *Q* factor of the system at the bonding mode. These results can be used to improve the nonlinear optical efficiency of metamaterials. The dependence relationship of the *Q* factor on the distance between the SRRs was also investigated, and this relationship provides a method for designing a nanocavity with an adjustable *Q* factor.

Recently, the coupling mechanism between two stacked SRRs was found not only to be determined by the distance between the two elements, but also to depend on the relative twist angle, $\varphi$ [41]. In another work by Giessen and colleagues, the hybridization effect of magnetic resonance was also observed in four stacked SRRs [42]. In addition to in identical resonators, hybrid modes were found in coupled structures composed of different resonators, including SRR pairs [43], cut-wire pairs [44], tri-rods [45, 46], and nanosandwiches in defective photonic crystals [47]. These hybrid modes could lead to some new interesting and useful properties, such as optical activity [38, 39] and omni-directional broadband negative refraction [45].

**III. Coherent collective waves in one-dimensional meta-chains**

In the above section, we presented the hybridization effect of hybrid modes among several coupled resonators. In this section, we will generalize the theoretical model to one-dimensional meta-chains of coupled magnetic resonators. We will show that the collective excitation of infinite magnetic atoms in metamaterials can induce a new kind of coherent collective waves, namely, an magnetic plasmon (MP) wave.

Linear chains of closely spaced metal nanoparticles have been intensely studied in recent years. Due to the strong near-field coupling interaction among these nanoparticles, a coupled electric plasmon propagation mode can be established in this chain and can be used to transport EM energy in a transverse dimension that is considerably smaller than the corresponding wavelength of illumination [48-53]. As this system can overcome the diffractive limit, it can function as a novel kind of integrated sub-wavelength waveguide.

According to the classic electrodynamics theory, the radiation loss of a magnetic dipole is substantially lower than the radiation of an electric dipole of a similar size [54]. Thus, using MP to guide EM energy over long distances has great potential for direct application in novel sub-diffraction-limited transmission lines without significant radiation losses.

Indeed, MP resonance has been already applied to a one-dimensional sub-wavelength waveguide in the microwave range [55-57]. Shamonina et al. proposed a propagation of waves supported by capacitively loaded loops using a circuit model in which each loop is coupled magnetically to a number of other loops [55]. Since the coupling is due to induced voltages, the waves are referred to as magneto-inductive waves (MI). MI waves propagating on such 1-D lines may exhibit both forward and backward waves depending on whether the loops are arranged in an axial or planar configuration. Moreover, the band broadening could be obtained due to the excitation of MI waves, and the bandwidth changes dramatically as we vary the coupling coefficient between the resonators [58]. A kind of polariton mode could be formed by the interaction between the electromagnetic and MI waves, resulting in a tenability of the range where $\mu$ becomes negative [57]. In a biperiodic chain of magnetic resonators, the dispersion of the MI wave will be split into two branches analogous to acoustic waves in solids, and this can be used to obtain specified dispersion properties [59, 60]. In addition to this kind of MI wave, electro-inductive (EI) waves were also reported in the microwave range [61]. Further, the coupling may be either of the magnetic or electric type, depending on the relative orientation of the resonators. This causes the coupling constant between resonators to become complex and leads to even more complicated dispersion [62]. Up to now, a series of microwave devices based on MI waves have been proposed, such as magneto-inductive waveguides [63], broadband phase shifters [64], parametric amplifiers [65], and pixel-to-pixel sub-wavelength imagers [66, 67].

However, in the optical range, the ohmic loss inside metallic structures is much higher than in the microwave range. The MI coupling between the elements is not strong enough to transfer the energy efficiently. In order to improve the properties of the guided MP wave, the exchange current interaction between two connected SRRs is proposed [68], which is much stronger than the corresponding MI coupling. Figure 4 (b) shows one infinite chain of SRRs constructed by connecting the unit elements (see Figure 4 (a)) one by one. The magnetic dipole model can be applied to investigate this structure. If a magnetic dipole, $\mu_m$, is assigned to each resonator and only nearest neighbor interactions are considered, then the Lagrangian and the dissipation function of the system can be written as

$$\mathcal{L} = \sum_m \left( \frac{1}{2} L \dot{q}_m^2 - \frac{1}{4C}(q_m - q_{m+1})^2 + M \dot{q}_m \dot{q}_{m+1} \right)$$
$$\mathfrak{R} = \sum_m \frac{1}{2} \gamma \dot{q}_m^2$$

(4)

Substitution of Equation (4) in the Euler-Lagrangian equations yields the equations of motion for the magnetic dipoles

$$\ddot{\mu}_m + \omega_0^2 \mu_m + \Gamma \dot{\mu}_m = \frac{1}{2} \kappa_1 \omega_0^2 (\mu_{m-1} + 2\mu_m + \mu_{m+1}) - \kappa_2 (\ddot{\mu}_{m-1} + \ddot{\mu}_{m+1}) \quad (5)$$

where $\kappa_1$ and $\kappa_2$ are the coefficients of the exchange current and MI interactions respectively. The general solution of Equation (5) corresponds to an attenuated MP wave: $\mu_m = \mu_0 \exp(-m\alpha d) \exp(i\omega t - imkd)$, where $\omega$ and $k$ are the angular frequency and wave vector, respectively, $\alpha$ is the attenuation per unit length, and d is the size of the SRR. By substituting $\mu_m(t)$ into Equation (5) and working in a small damping approximation ($\alpha d \ll 1$), simplified relationships for the MP dispersion and attenuation are obtained:

$$\omega^2(k) = \omega_0^2 \frac{1 - \kappa_1[1 + \cos(kd)]}{1 + 2\kappa_2 \cos(kd)} \quad (6)$$

In Figure 4 (e), numerically and analytically estimated MP dispersion properties are depicted as dots and solid curves respectively. In contrast to the electric plasmon modes in a linear chain of nanosized metal particles where both transverse and longitudinal modes can exist, the magnetic plasmon is exclusively a transversal wave. It is manifested by a single dispersion curve that covers a broad frequency range, $\omega \in (0, \omega_c)$, with the cutoff frequency $\hbar \omega_c \approx 0.4eV$. Finally, it is important to mention that the MP properties can be tuned by changing the material used and the size and shape of the individual SRRs.

Magnetic resonance coupling between connected split ring resonators (SRRs) and magnetic plasmon (MP) excitations in another type of connected SRR chains has also been studied [69]. By changing the connection configuration (see Figure 5 (a) and (b)), the chain provides two kinds of MP bands formed by the collective magnetic resonances in the SRRs. From the extracted dispersion properties of MPs, forward and backward characteristics of the guided waves are well exhibited corresponding to the homo- and hetero-connected chains, which is shown in Figure 5 (c) and (d). Due to the conductive coupling the revealed MP waves both have wide bandwidth even starting from the zero frequency. These results are suggested to provide instructions to build new kinds of subwavelength waveguides.

Generally, the MP resonance frequency increases linearly with the decrease in the overall SRR size. However, the saturation of the magnetic response of the SRR at high frequencies prevents this structure from achieving high-frequency operation [70]. In addition, the complicated shape and narrow gap of the SRRs make experiments very challenging. The slit-hole resonator (SHR) [71] is considered as a good alternative to make sub-wavelength waveguides because of their simple structures and high working-frequency regime. Figure 6 (a) presents the geometry model of an SHR structure. The corresponding LC transmission line model is shown in Figure 6 (b). We also fabricate a sample of a diatomic SHR structure with an FIB (focused ion beam) method, whose image is presented in Figure 6 (c). In the bi-atomic SHR structure, there exist two kinds of magnetic resonance, the bonding mode and the anti-bonding mode. Therefore, in the bi-atomic chain of SHR, these two modes expanded into two bands, acoustic modes and optical modes, which is shown in Figure 6 (d) and (e). Finally, the measured transmittance at different incident angle and different frequency is plotted in Figure 6 (g), which shows an evident MP mode band that is consistent with the optical branch in the theoretical results.

In the category of magnetic plasmon structures, the nanosandwich structure is also a good choice to reach high resonant frequency, even to light frequency region. Figure 7 (a) presents the geometry of a single nanosandwich, composed of two metallic nanodiscs and a dielectric middle layer [72]. The anti-parallel currents in the metallic slabs induce a high intensity and confined magnetic field at a certain frequency, thus it can be seen as a magnetic atom. The frequency spectrum and field distribution of such nanosandwich is shown in (b)-(d), respectively. Such a magnetic atom can be used to construct a linear magnetic chain. Due to the near-field electric and magnetic coupling interactions, an MP propagation mode is established in this 1-D system. When excited by an EM wave, a strong local magnetic field is obtained in the middle layer at a specific frequency (Figure 7 (e)). For this magnetic plasmon resonance mode, the corresponding electric fields are given in Figure 7 (f). It should also be mentioned that such an MP waveguide is subwavelength, whose energy flow cross section is plotted in (g). The field is confined in a small area smaller than the wavelength scale. Through a Fourier transform method, the wave vectors of this MP wave at different EM wave frequencies are calculated. Then, the MP wave's dispersion property is obtained (shown as a white line in Figure 7 (h)). The light line in free space is also given as the black dotted line in the figure. The MP curve is divided into two parts by the light line. The part above the light line corresponds to bright MP modes whose energy can be radiated out from the chain, while the part below the light line corresponds to dark MP modes whose energy can be well confined within the chain. It is easy to see that the bright MP modes are much weaker than the dark MP modes for their leaky property. Therefore, only those EM waves in the frequency range of the dark MP modes can be transferred efficiently without radiation loss.

We also do some structure engineering on the nanosandwich chain. Once the above results for a mono-periodic chain of nanosandwiches have been generalized to graded structures [73], some new interesting properties, such as slow group velocity and new type of field distribution, are found in these more complex structures. The dispersion relation of the graded chain is shown in Figure 8 (a). The MP modes can be divided into three parts, gradon (the special mode belonging to the graded structure), extended mode, evanescent mode. The field distributions of these three types are quite different and the location the field of gradon is strongly depended on the frequency, which is presented in Figure 8 (b). By employing this property, one can manage a wavelength selective switch. Three ports switch and four ports switch can be realized in this graded nanosandwich chain. The field distributions of the magnetic field corresponding to different mode of these switches are presented in Figure 8 (c). Some new interesting properties, such as slow group velocity and band folding of MP waves, are found in these more complex structures.

**VI. Excitation of coupled plasmon modes in two-dimensional meta-crystals**

In addition to the abovementioned 1-D structures, the MP mode introduced by the coupling effect in 2-D systems is also an interesting topic. For 2-D metamaterials, the most important applications are negative refraction, focusing, and superlensing. How do the coupling interactions between elements affect the above processes? They cannot be handled by the conventional effective medium theory.

In the microwave range, MI wave theory has been proposed to deal with the coupling effect in the 2-D system. An MI superlens was proposed based on employing the coupling between resonators [66, 67], which eliminates the weakness of Wiltshire's first Swiss-roll superlens [12, 13] and has potential for MRI applications. The focusing of indefinite media, originally treated by Smith [74], has been investigated by Kozyrev with the aid of MI wave theory. It was found that partial focusing and multiple transmitted beams can be formed by the excitation of MI waves [75]. A further comparison between effective medium theory and MI wave theory was also given by Shadrivov [58], in which the reflection and refraction of MI waves on the boundary of two different effective media were studied. It was shown that both positive and negative refraction may occur under some configurations of the elements [76]. Another interesting finding is that spatial resonances could be formed by the propagation of MI waves on a 2-D array of magnetic resonators [77]. Different boundary conditions will produce different current and magnetic field distributions.

In the optical frequency range, fishnet structures are well-known magnetic metamaterials [20], constructed by a metal/insulator/metal (MIM) sandwich with perforated periodic nanohole arrays, schematically shown in Figure 9. This structure can somewhat be regarded as an extension of the metal film perforated with nanoholes that exhibits well-known extraordinary optical transmission (EOT) discovered by T. W. Ebessen in 1998 [78]. However, the most significance of this fishnet structure is owing to its novel negative refraction property revealed by S. Zhang in 2005 [20]. The fundamental physics to realize the negative index in this structure is based on the artificial "magnetic atoms" consisted of the LC-resonance between the two coupled metallic segments and the "electric atoms" from continuous metallic strips parts, which produce simultaneous negative effective permeability and permittivity correspondingly. Based on such kind of design, the performance of the negative index property was subsequently improved by the means of structural optimizations [79-81]. Simultaneous negative phase and group velocity of the light in this fishnet metamaterial was obtained via the method of interference [82]. Even more, by further elaborate minimization, the negative index mode was pushed to red color wavelength [83].

On the other hand, some attempts were carried out in considering the effect by stacking multilayer fishnet structures. It was numerically shown that the optical transmission loss for this negative refraction mode can even lower down by stacking more fishnet elements in the propagation direction [84]. However, the underlying physics remains obscure. Focusing on this point, our group made a detailed investigation on the coupling effect not only in the stacked fishnets but also extending the original double-metal-layer (DML) fishnet to three-metal-layer (TML) one and even two more multilayers [85], in which coupling effect may be more remarkable. Two distinct modes were found in this TML structure (denoted as R1 and R2), which is actually due to the coupling of two magnetic resonances, shown in Figure 10 (a-b). This coupling effect was subsequently explained by a coupled circuit model associated with a metallic skin depth related layer thickness. The improved transmission property can be realized at a certain range of metal layer thickness as shown in Figure 10 (c), indicating a reduced loss in such a coupled system. This result in a certain extent explained the former results [84], and was later proved by our experimental result. Moreover, this coupling mechanism provides an encourage vision to realizing a real 3-D negative index metamaterial, which is ultimately fulfilled by X. Zhang group in 2008 [22]. They fabricated a prism consisted of multilayered fishnets via the focus ion beam etching, and the negative refraction phenomena was definitely revealed by evaluating the position of the transmitted light beam at a particular wavelength.

It has been revealed that the coupling effect plays an important role in improvement of NIM performance. Actually, the coupling effects are still presented in the in-plane dimensions. G. Dolling et al. noticed this effect and they found magnetization waves in such fishnet structure in end of 2006 [86]. However, what they observed is only a lowest mode of a kind of collective magnetic excitation due to the in-plane coupling, which was later called as magnetic plasmon polaritons (MPPs). Actually, our group made a detailed investigation on the plasmonic modes of the artificial "magnetic atoms". Multiple MPP modes associated with reciprocal vectors of the lattice was convincingly demonstrated [87]. Figure 11 (a-b) show the magnetic field distributions of two mentioned MPP modes associated with reciprocal vectors of G(0,1) and G(1,1). Figure 11 (c) is a calculated transmission map with the $SiO_2$ layer thickness ranging from 25 nm to 55 nm, where two MPP modes and SPP modes are clearly exhibited. Using this plasmonic property, we also can easily modulate the eigen frequency of the MPP modes via adjusting the structure parameters, as Figure 11 (d-e) shows.

Afterwards, the dispersion properties of the MPP modes in fishnet structure with rectangular hole array was studied in succession [88]. By carefully investigation on the transmission property on the oblique incidence for the *s*- and *p*-polarization cases, we found a polarization dependent dispersion property of the concerned MPP modes, which was indicated in transmission maps for these two polarizations, as shown in Figure 12. From that, we can see that the MPP(1,1) mode in s-polarization actually exhibits much larger dispersions than the lowest mode, resulting in two split modes MPP(+1,1) and MPP(-1,1) with the degeneration broken up. This is very similar as the property of SPP. As for the anisotropic property, we contributed them to different coupling intensities among the artificial "magnetic atoms" with the 2D plane. Ultimately, we used a formula to describe the dispersion property of MPPs in the fishnet structure as

$$\lambda_{MPP(m,n)} = \frac{2\pi c}{|(\delta_y \cdot k_x \pm mG_x)\hat{x} + (\delta_x \cdot k_y \pm nG_y)\hat{y}|}\left(\frac{1}{\omega_{LC}}\right) \quad (7)$$

where $\delta_x$ and $\delta_y$ are introduced to describe the polarization. When the incident light is *x*-polarized (*y*-polarized), $\delta_x=1$ and $\delta_y=0$ ($\delta_y=1$ and $\delta_x=0$).

Recently, optical transmission property in this fishnet metamaterial attracts more attentions due the discussions about the origin of the magnetic response. Garcia-Vidal group proposed a relative general theory to explain the negative refractive response in this system [89]. They considered that the electrical response of these structures is dominated by the cutoff frequency of the hole waveguide whereas the resonant magnetic response is due to the excitation of gap surface plasmon polaritons (gap-SPP) propagating along the dielectric slab. Furthermore, they predicted that the negative index mode is dispersive with the parallel momentum of the incident light. This conclusion is somewhat consistent with our experimental results [88], although further particular consideration may be still supplied theoretically concerning on the different polarizations. Actually, our description in the term of MPP may not conflict with the explanation of the gap-SPP, because the magnetic resonance is identical with the antisymmetric electric resonance mode in a coupled (MIM) structure. Thereafter, the explanation of gap-SPP on the negative refractive response in fishnet metamaterial was further confirmed by more detailed analysis [90-92]. Ortuno, et al provided an elaborate interpretation on the role of surface plasmon in the transmission properties, and came out the contributions of coupled external SPP and internal SPP [90], which is privately considered as another kind of description of the artificial electromagnetic excitations with respect to our findings about the optical magnetism induced by SPP excitations [93]. Actually, our own recent studies show that the MPP mode described in [87, 88] is surely identical to the gap-SPPs due to the same lattice modulation, but only from different point of view. Till now, most of researchers have come out the consensus that multiple magnetic modes presented in the fishnet metamaterial with negative refractive response and the underlying physics is due to the internal coupled SPP excitations. From the point, it is still a coupling effect that has aroused so many interests and discussions.

To give a short summary, the coupling effect endows the fishnet structure with fruitful interesting properties and the complicated mechanism was recognized more and more clearly. In the following, this novel structure in quasi 2-D system may not only exhibits interesting properties in the far field optics, but also be expected with fancy features in near field optical modulations and integrate applications.

**V. Outlook**

Although metamaterials composed of uncoupled magnetic resonance elements have been successfully applied to produce intriguing effects such as negative refraction, cloaking, and superlensing, all of these were devised within a very narrow frequency range around a specific resonance frequency. This disadvantage restricts the practical applications of metamaterials.

In addition to the abovementioned applications of the linear optical effect, due to the great enhancement of the local field inside magnetic resonators, magnetic metamaterials have also been proposed for use in nonlinear optical processes, such as SERS [5], SHG [94, 95], nanolasers [96, 97], and SPASER [98-100]. However, the nonlinear optical processes that occur between waves of several different frequencies typically require a broad frequency bandwidth. The narrow single resonance property of conventional metamaterials is a considerable disadvantage for their potential nonlinear optical applications.

The MP modes introduced by the coupling effect in metamaterials may provide a possible way to overcome the abovementioned obstacle. As described in this paper, hybrid MP resonance modes could be attained in a system with several coupled resonators. This hybridization effect results in multiple discrete resonance frequencies of magnetic metamaterials. When all the resonance elements in metamaterials are coupled together through a particular method, the multiple resonance levels will be extended to a continuous frequency band. Therefore, the excitation of MP modes in such metamaterials can be continually tuned within a rather wide range. Compared with conventional metamaterials made from uncoupled elements, this kind of broadband tunable magnetic metamaterial based on the coupling effect will have much more interesting and prospective applications, especially on the nonlinear optical effect. Based on this discussion, we anticipate that many novel metamaterials and nanophotonic devices will be developed from coupled resonator systems.

**Acknowledgements** This work is supported by the National Natural Science Foundation of China (No. 10704036, No. 10874081, No. 60907009, No. 10904012, No. 10974090, and No. 60990320), and by the National Key Projects for Basic Researches of China (No.2009CB930501, No.2006CB921804 and No. 2010CB630703).

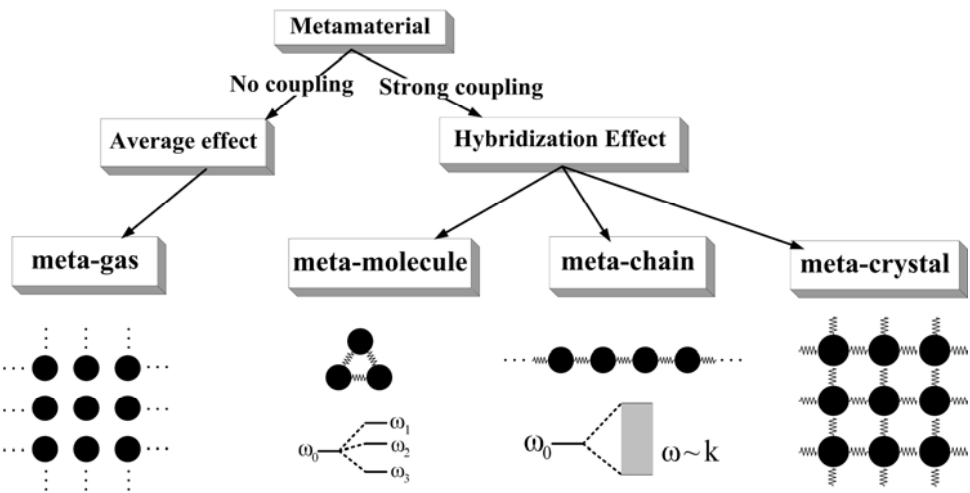

**Figure 1** "average effect" and "hybridization effect" in metamaterials

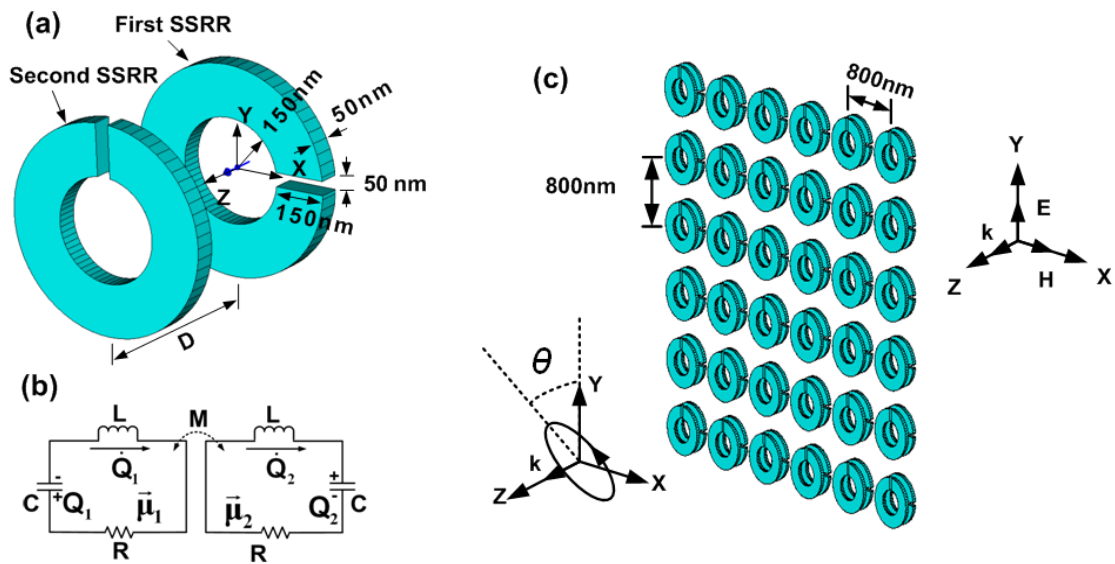

**Figure 2** (a) Structure of a magnetic dimer; (b) equivalent LC circuit; (c) metamaterial made of identical dimer elements.

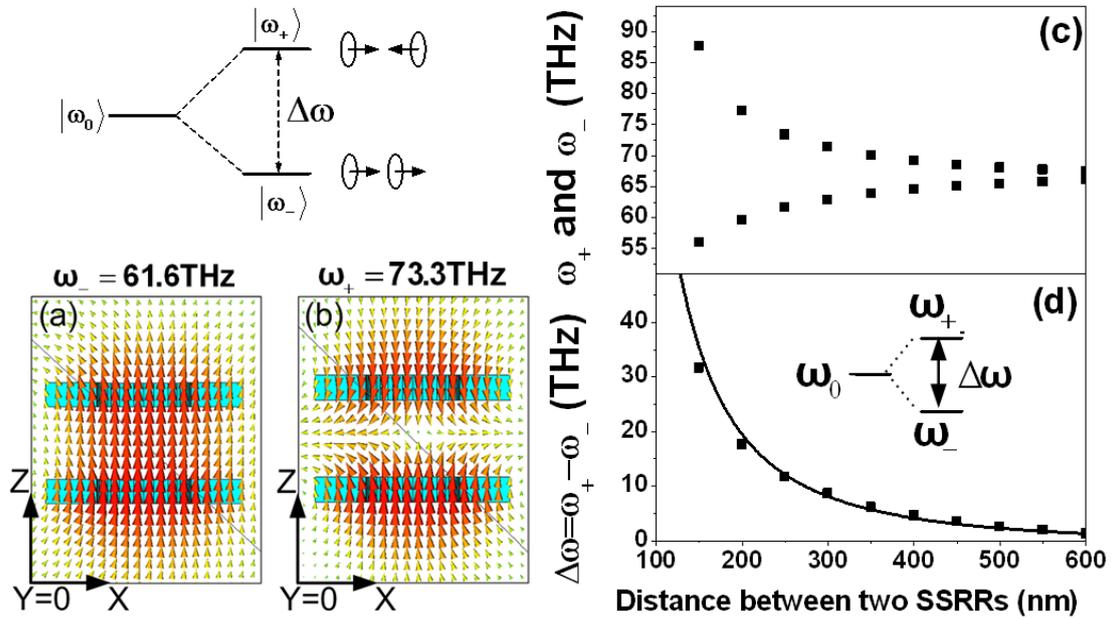

**Figure 3** The local magnetic field profiles for the (a) bonding and (b) antibonding MP modes; The dependence of the resonance frequencies (c) and the frequency gap (d) on the distance between two SRRs.

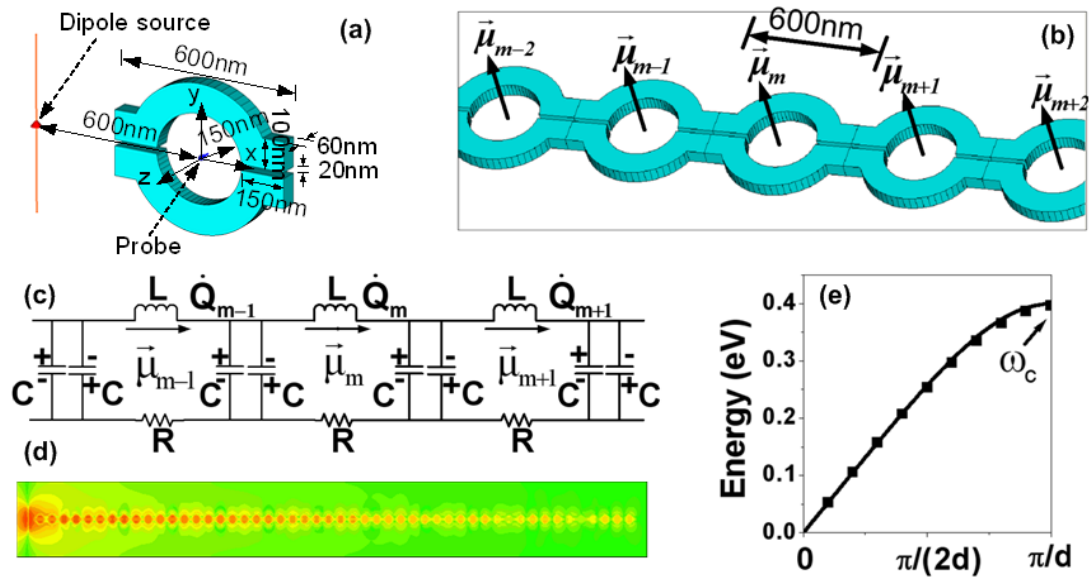

**Figure 4** (a) Structure of a single SRR; (b) one-dimensional chain of SRRs; (c) equivalent circuit of the chain; (d) FDTD simulation of MP wave propagation along the chain; (e) dispersion curve of the MP wave.

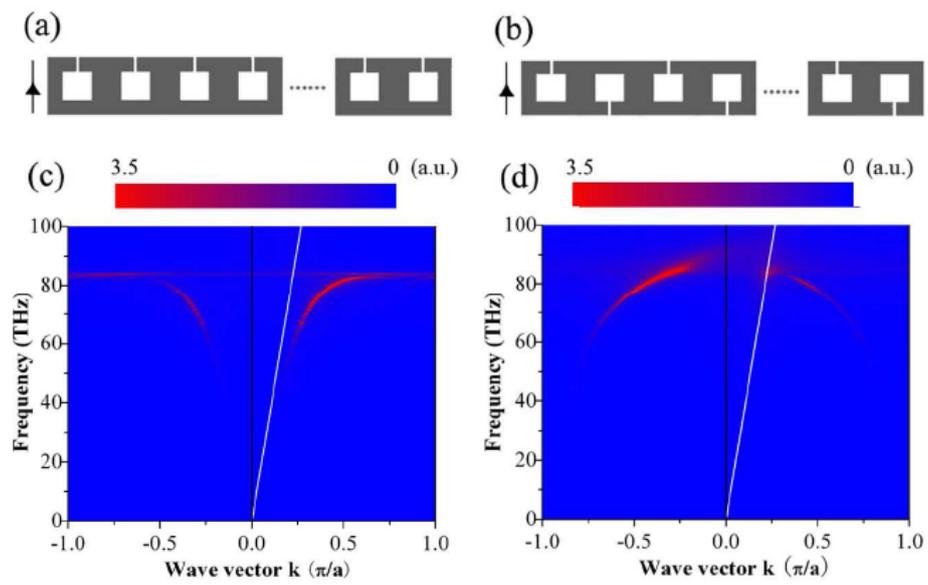

**Figure 5** Subwavelength waveguides constituted by the SRR chains with (a) homo-connection and (b) anti-connection. (c) and (d) are the Fourier Transformation map in the ω-$k$ space corresponding to the waveguides (a) and (b), respectively.

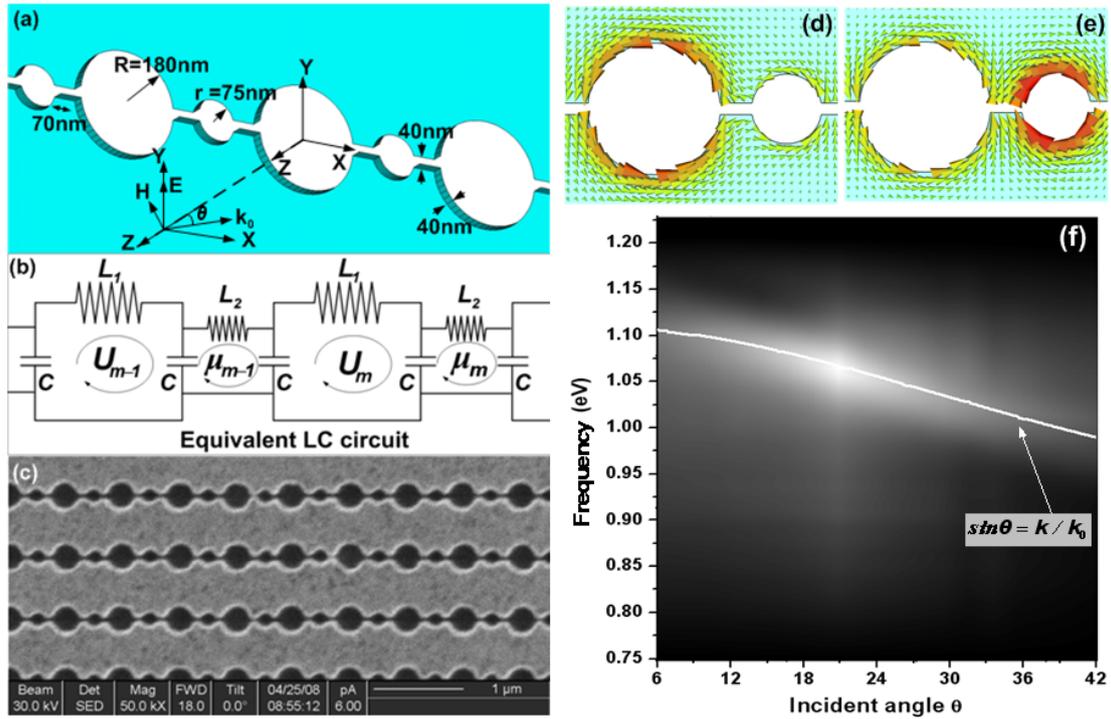

**Figure 6** (a) Structure of a diatomic chain of SHRs; (b) Equivalent *LC* circuit of the chain; (c) FIB image of the fabricated sample. The local current distributions are calculated for the (d) optical and (e) acoustic modes; (f) Measured transmission map and the calculated angular dependence curve of the optical MP mode.

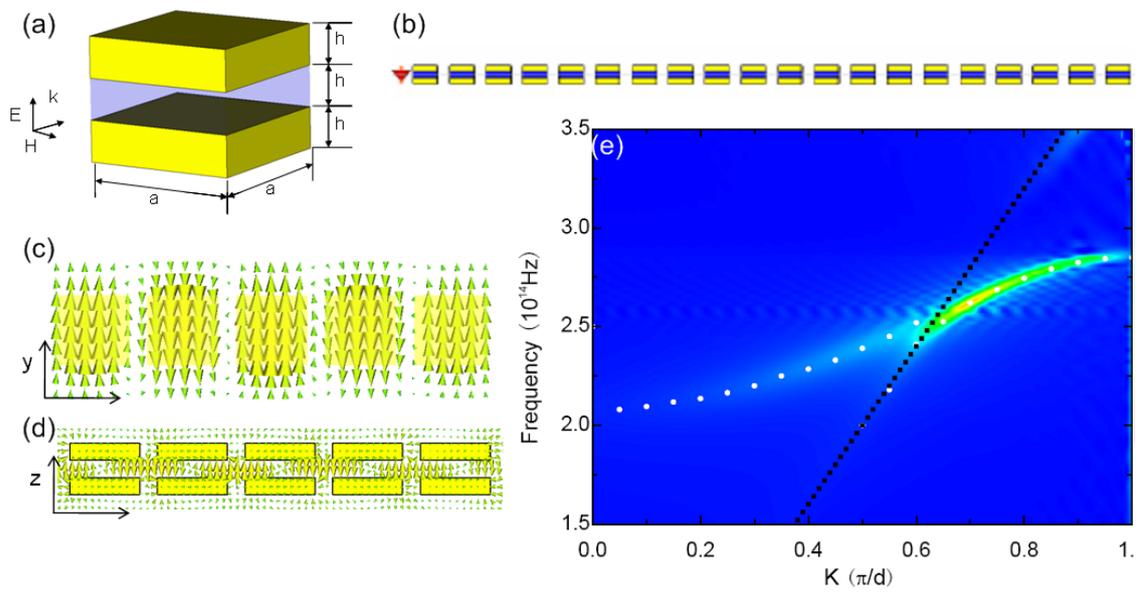

**Figure 7** (a) Structure of a single nanosandwich; (b) one-dimensional chain of nanosandwich; (c) electric field and (d) magnetic field of the MP wave propagation along the chain; (e) dispersion curve of the MP wave.

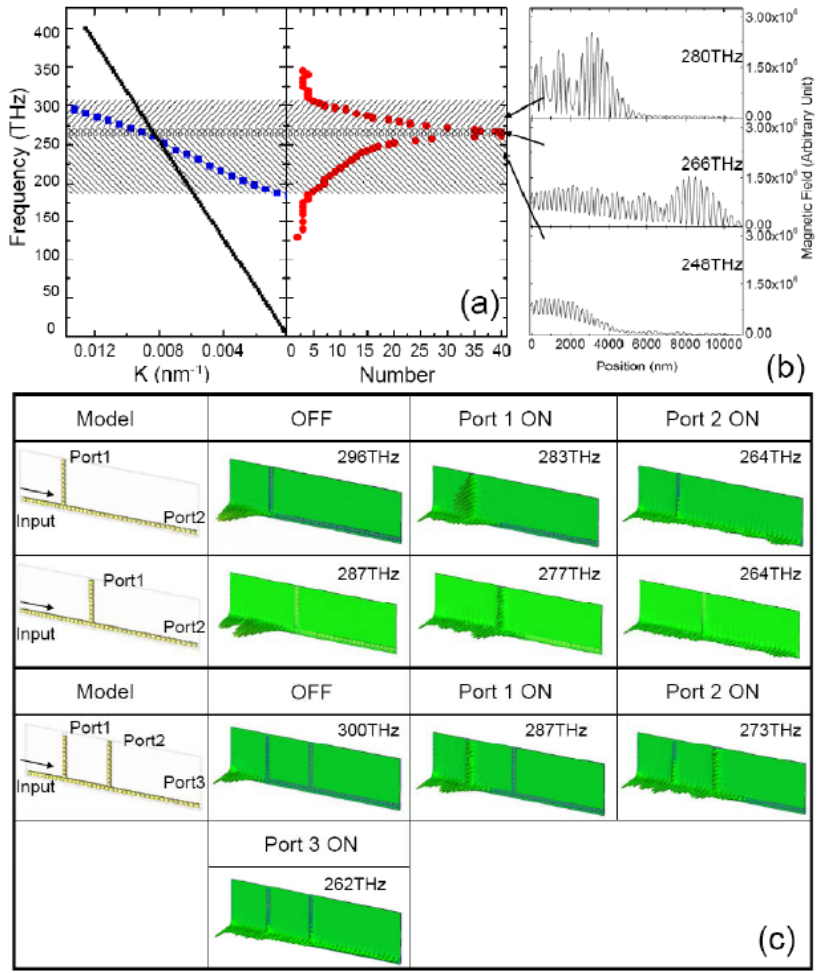

**Figure 8** (a) Dispersion relation and propagating length of the graded nanosandwich chain; (b) Filed distributions of three different modes; (c) The field distributions corresponding to different modes in both three port switch and four port switch.

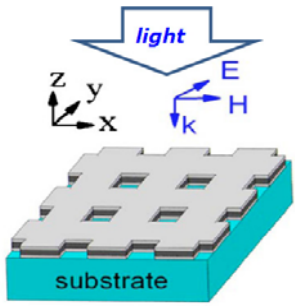

**Figure 9** Schematics of fishnet metamaterial.

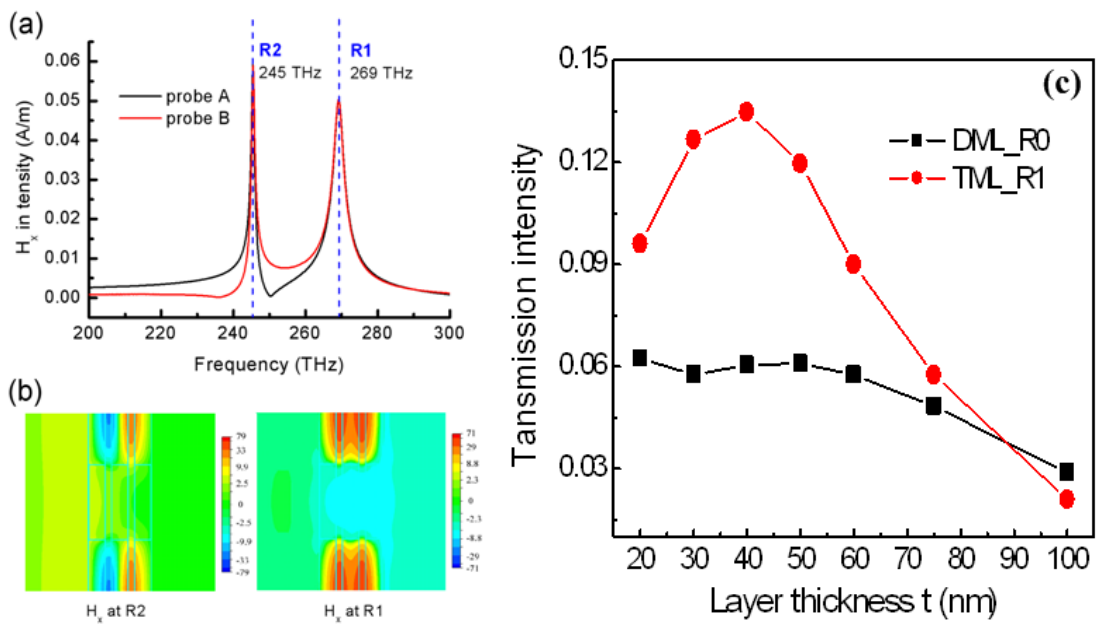

**Figure 10** (a) Detected $H_x$ intensities of TML structure with the frequency ranging from 200 to 300 THz by two probes, which are located at the neighboring layer gaps. (b) The corresponding simulated $H_x$ distributions at the frequencies of R1 and R2. (c) Transmission intensities of mode R0 and R1 versus the layer thickness for the DML and TML structures, respectively.

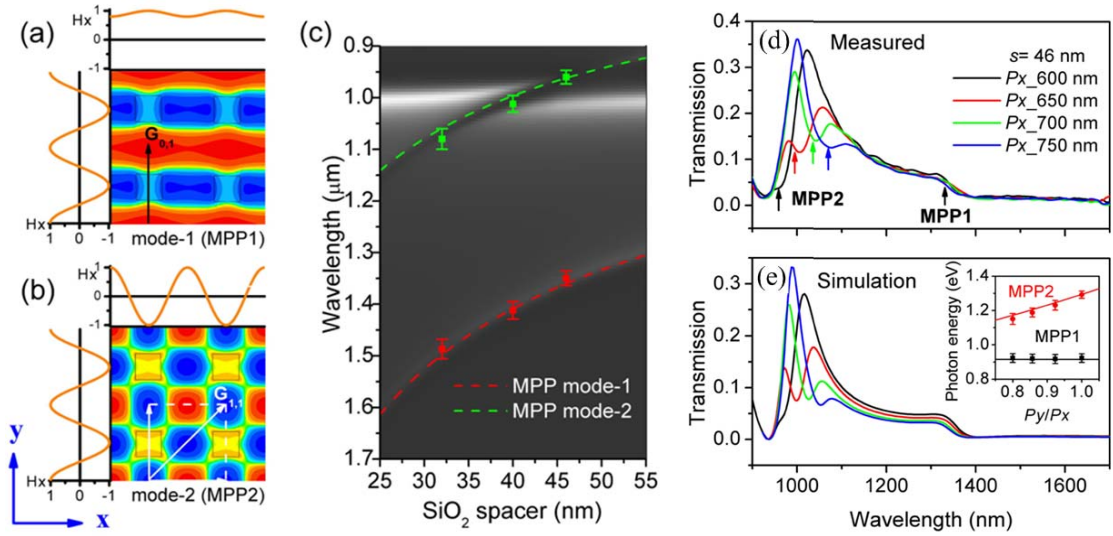

**Figure 11** (a) Simulated magnetic field distribution map for mode 1 (a) and for mode 2 (b), in xy plane (z=0) in the center of the middle SiO$_2$ layer. (c) Calculated transmittance map of samples as middle layer thickness s ranging from 25 to 55 nm, where the MPP modes and SPP mode can be clearly observed. Measured (d) and calculated (e) transmission spectra of the fishnet samples with various period of the hole lattice in x direction.

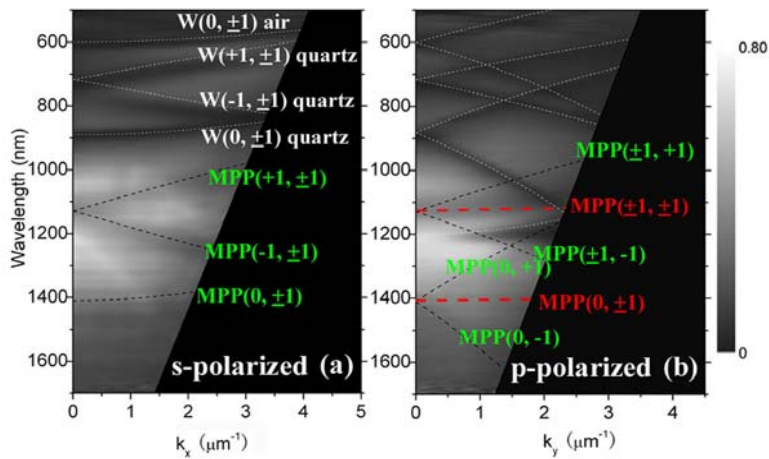

**Figure 12** Measured transmission maps on an intuitive gray scale versus wavelength and vavevectors for (a) s-polarization and (b) p-polarization.